\documentclass
[reprint,prl,a4paper,amsfonts,amssymb,superscriptaddress,onecolumn]{revtex4}
\usepackage{amssymb}
\usepackage{amsfonts}
\usepackage{amsmath}
\usepackage{graphicx}
\usepackage{array}
\usepackage{SIunits}
\usepackage{ulem}
\allowdisplaybreaks[3]

\begin{document}

\title{Tunnel junctions with moir\'{e} superlattice as barrier}
\author{Henan Fang}
\affiliation{College of Electronic and Optical Engineering, Nanjing University of Posts and Telecommunications, Nanjing 210023, China}
\author{Mingwen Xiao\footnote{Email: xmw@nju.edu.cn}}
\affiliation{Department of Physics, Nanjing University, Nanjing 210093, China}

\begin{abstract}
Recently, moir\'{e} superlattices have attracted considerable attentions because they are found to exhibit intriguing electronic phenomena of tunable Mott insulators and unconventional superconductivity. These phenomena are highly related to the physical mechanism of the interlayer coupling. However, up to now, there has not existed any theory that can completely interpret the experimental results of the interlayer conductance of moir\'{e} superlattice. In order to solve this problem, the superposition of periods and the corresponding coherence, which are the essential characteristics of moir\'{e} superlattice, should be considered more sufficiently. Therefore, it is quite necessary to introduce optical methods to study moir\'{e} superlattices. Here, we develop a theory for moir\'{e} superlattices which are founded on traditional optical scattering theory. The theory can interpret both the continuously decreasing background and the peak of the interlayer conductance observed in the experiments by a unified mechanism. We show that, the decreasing background of the interlayer conductance arises from the increasing strength of the interface potential, and the peak roots from the scattering resonance of the interface potential. The present work is crucial for understanding the interlayer coupling of the moir\'{e} superlattice, and provide a solid theoretical foundation for the application of moir\'{e} superlattice.

\noindent{\textbf{Keywords}: moir\'{e} superlattices, interlayer conductance, twist angle, moir\'{e} tunnel junctions, two-dimensional materials}
\end{abstract}

\maketitle

\section{Introduction}
The moir\'{e} phenomenon has a long history. It had already been used by the Chinese in ancient times for creating an effect of dynamic patterns in silk cloth \cite{rf1}. For modern scientific research into the moir\'{e} phenomenon, it started only in the second half of the 19th century with pioneering work related to the diffraction-gratings \cite{rf1,rf2}. Until today, although the moir\'{e} phenomenon has been applied in many fields of science and technology such as strain analysis, the most key application is in the optical field \cite{rf1}. That is to say, the study of moir\'{e} phenomenon follows closely with the development of optics. Recently, the moir\'{e} phenomenon receives anew extensive attentions due to the new quantum phenomena discovered in the moir\'{e} superlattices \cite{rf3,rf4,rf5,rf6,rf7,rf8,rf9,rf10,rf11,rf12}, in particular, Cao et al. find that the moir\'{e} superlattices can lead to intriguing electronic phenomena of tunable Mott insulators \cite{rf13} and unconventional superconductivity \cite{rf14}.

In structure, a moir\'{e} superlattice is formed by vertically stacking two layered quasi-two-dimensional single crystals with a twist angle. The two quasi-two-dimensional crystals may be fabricated with the same or different materials, that is to say, a moir\'{e} superlattice can be homogeneous \cite{rf6,rf7,rf9,rf11,rf12,rf13,rf14,rf15,rf16,rf17,rf18,rf19,rf20,rf21,rf22,rf23,rf24,rf25,rf26,rf27,rf28,rf29,rf30,rf31} or heterogeneous \cite{rf4,rf5,rf32,rf33,rf34,rf35,rf36}. For homogeneous case, most of them are made up of graphite (graphene)/graphite (graphene) \cite{rf6,rf7,rf9,rf11,rf12,rf13,rf14,rf15,rf16,rf17,rf18,rf19,rf20,rf21,rf22,rf23,rf24,rf25,rf26,rf27,rf28,rf29}. As far as we know, the familiar materials of the quasi-two-dimensional crystals used to form moir\'{e} superlattices include graphite (graphene), MoS$_{2}$, black phosphorus, WSe$_{2}$ and hexagonal boron nitride. Structurally, they are all of hexagonally layered materials.

Evidently, in order to clarify the physical mechanism for those new quantum phenomena and to improve the performance of the electronic devices based on moir\'{e} superlattices, it is significant and crucial to obtain a comprehensive understanding of the relationship between the interlayer conductance and the twist angle \cite{rf37}.

Experimentally, there are a few researches which have studied the relationship between the interlayer  conductance and the twist angle \cite{rf37,rf38,rf39,rf40,rf41,rf42}. The common feature in those experiments is that the interlayer  conductance decreases initially with the twist angle within the domain $[0^{\circ},\,30^{\circ}]$, and then exhibits a $30^{\circ}$ reflection symmetry and a $60^{\circ}$ periodicity, which arise from the hexagonal symmetry of the layered quasi-two-dimensional single crystals. Furthermore, and most importantly, Refs. \cite{rf38} and \cite{rf39} find that there are two sharp conductivity peaks around the twist angles of $21.8^{\circ}$ and $38.2^{\circ}$ (the peak around $38.2^{\circ}$ is the mirror peak of the one around $21.8^{\circ}$, according to the $30^{\circ}$ reflection symmetry). To sum up, the experiments show that, with the variation of the twist angle, the interlayer  conductance behaves like a continuously decreasing background plus one sharp peak within the domain of $[0^{\circ},\,30^{\circ}]$, apart from a $30^{\circ}$ reflection symmetry and a $60^{\circ}$ periodicity due to the crystal symmetry.

Theoretically, there are two models for the background in the previous literatures: The first is the phonon-assisted transport model introduced by Refs. \cite{rf37}, \cite{rf38}, \cite{rf41} and \cite{rf42}; the second is the density functional theory simulation introduced by Ref. \cite{rf40}. The former agrees well with the experiments for large twist angles, but it significantly violates the experiments for small twist angles \cite{rf38}. The latter can interpret the background in MoS$_{2}$/graphene moir\'{e} superlattices, however, it can only be applicable to the MoS$_{2}$/graphene case, but not work for other moir\'{e} superlattices, such as graphite(graphene)/graphite(graphene) moir\'{e} superlattices \cite{rf40}. In addition, the two models can at most produce the continuous background, neither of them can work out the peaks present in the interlayer conductance \cite{rf37,rf38,rf40,rf41,rf42}. On the other hand, Ref. [43] predicted theoretically that there should emerge conductivity peaks at a series of commensurate angles: $13.2^{\circ}$, $21.8^{\circ}$, $27.8^{\circ}$, $32.2^{\circ}$, $38.2^{\circ}$ and $46.8^{\circ}$. Subsequent experiments do observe two peaks \cite{rf38,rf39}. It is found that they correspond to the commensurate angles  $21.8^{\circ}$ and $38.2^{\circ}$, respectively. All the other four predicted peaks, which correspond respectively to the commensurate angles $13.2^{\circ}$, $27.8^{\circ}$, $32.2^{\circ}$ and $46.8^{\circ}$, are regrettably absent in the experiments. Another regret for the theory of Ref. \cite{rf43} is that it can only make the peaks, but can not describe the background. In a word, neither the background nor the peaks has been well explained theoretically. Needless to say, the background and the peaks have not been explained by a unified mechanism as yet.

That motivates us to endeavor in this paper to develop a microscopic theory to interpret both the continuous background and the sharp peaks present in the interlayer  conductance by using a unified mechanism.

Previously, a spintronic theory for the magnetic tunnel junctions (MTJs) with single-crystal barrier has been developed by us \cite{rf44,rf45,rf46,rf47}. The theory is founded on the traditional optical scattering theory \cite{rf48}. Within it, the barrier is treated as a diffraction grating with intralayer periodicity. It is found that the periodic grating can bring strong coherence to the tunneling electrons. This coherence is the main mechanism responsible for the MTJs with single-crystal barrier. So far, the theory has successfully explained the basic properties of the MgO-based MTJs \cite{rf44,rf45,rf46,rf47}.

Physically, a moir\'{e} superlattice is nearly insulating along the interlayer, or longitudinal, direction \cite{rf49}, therefore, when discussing the longitudinal transportation of electrons, it can be regarded as a tunneling barrier, and obviously, the so-called interlayer conductance is a kind of tunneling conductance \cite{rf40}. Besides, as stated above, every periodic atomic layer within a moir\'{e} superlattice can be considered as a diffraction-grating in optics \cite{rf44,rf45,rf46,rf47}. Therefore, a moir\'{e} superlattice is essentially in physics the same as a single-crystal tunneling barrier, the previous tunneling theory for MTJs \cite{rf44,rf45,rf46,rf47} can be transplanted to handle moir\'{e} superlattices. That constitutes the main object of this paper. Clearly, the transplanted theory belongs to optics in methods, which is in harmony with the traditional theories in the moir\'{e} field \cite{rf1}. As will be seen in the following, it can describe both the continuous background and the sharp peaks present in the interlayer  conductance by a unified mechanism. Furthermore, we would like to propose a new type of electronic devices, i.e. moir\'{e} tunnel junctions. Based on the present work, their working mechanism will be analyzed.

\section{Model and Methods}

Now, let us consider a tunneling junction with a moir\'{e} superlattice as the barrier, which is sketched diagrammatically in Figs. 1(a), 1(b) and 1(c) where the two electrodes are common metals, the $d_{1}$ and $d_{2}$ are, respectively, the thicknesses of the upper and lower layered single crystals of the moir\'{e} superlattice, and $\alpha$ the twist angle between the two layered single crystals. For convenience, we shall suppose that the tunneling electrons first pass through the upper layered single crystal with the thickness of $d_{1}$, and subsequently pass through the lower layered single crystal with the thickness of $d_{2}$. As depicted in Figs. 1(a) and 1(b), there exists an interface between the upper and lower layered single crystals, physically, it represents the deformation from the lattice twisting. As a result, the barrier potential will include three parts: the periodic potential $U_{1}$ arising from the upper layered single crystal, the periodic potential $U_{2}$ arising from the lower layered single crystal, and the potential $V$ arising from the interface of twisted deformation. For simplicity, we only discuss here the case that the two layered single crystals are fabricated with the same materials, i.e., the moir\'{e} superlattice is homogeneous. The discussion can be easily generalized to heterogeneous moir\'{e} superlattices.

\begin{figure}[!ht]
\centering
\includegraphics[scale=0.8]{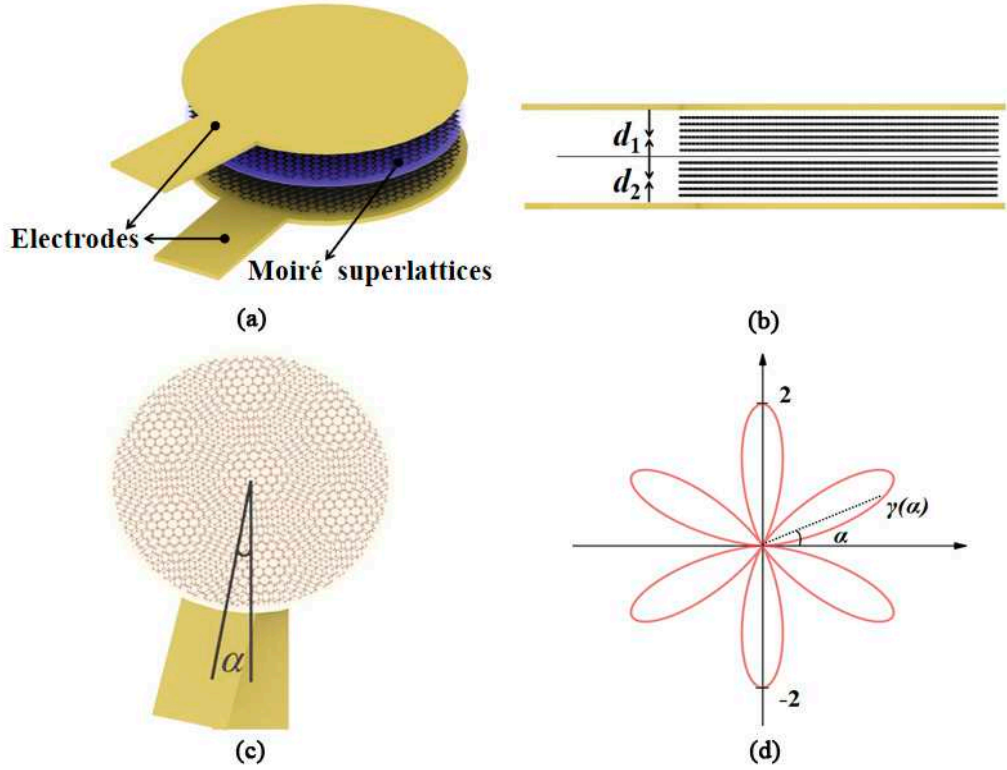}
\caption{The diagrammatic sketch of the studied system. (a) Schematic of the studied system. (b) Side view of the studied system. (c) Top view of the studied system. (d) The strength of the interface potential.}
\end{figure}

As such, the potentials $U_{1}$, $U_{2}$ and $V$ can be written as follows,
\begin{eqnarray}
  U_{1}(\mathbf{r}) &=& \sum_{l_{3}=0}^{m-1} \sum_{\mathbf{R}_{h}}
                       v(\mathbf{r}-\mathbf{R}_{h}-l_{3}\,\mathbf{a}_{3}) , \\
  U_{2}(\mathbf{r}) &=& \sum_{l'_{3}=0}^{n-1}\sum_{\mathbf{R}_{g}}
                        v(\mathbf{r}-d_{1}\mathbf{e}_{z}-\mathbf{R}_{g}-l'_{3}\,\mathbf{a}'_{3}),\\
  V(z) &=& -\gamma(\alpha)\delta(z-d_{1})(1-i\varepsilon). \label{Vgamma}
\end{eqnarray}
Here, $v(\mathbf{r})$ represents the atomic potential of the upper and lower single crystals of the barrier. The $m$ and $n$ denotes the total numbers of the atomic layers within the upper and lower single crystals of the moir\'{e} superlattice, respectively. The lattice vector $\mathbf{R}_{h} = l_{1}\, \mathbf{a}_{1} + l_{2}\, \mathbf{a}_{2}$, with $\mathbf{a}_{1}$ and $\mathbf{a}_{2}$ being the intralayer primitive vectors of the upper single crystal of the barrier, and $l_{1}$ and $l_{2}$ the corresponding integers; the lattice vector $\mathbf{R}_{g} = l'_{1}\, \mathbf{a}'_{1} + l'_{2}\, \mathbf{a}'_{2}$, with $\mathbf{a}'_{1}$ and $\mathbf{a}'_{2}$ being the intralayer primitive vectors of the lower single crystal of the barrier, and $l'_{1}$ and $l'_{2}$ the corresponding integers. Obviously, the $\alpha$ is the angle between the basis vectors $\mathbf{a}_{1}$ and $\mathbf{a}'_{1}$, or between $\mathbf{a}_{2}$ and $\mathbf{a}'_{2}$. As to $\mathbf{a}_{3}$ and $\mathbf{a}'_{3}$, they are the interlayer primitive vector of the upper and lower single crystals of the barrier, respectively, with $l_{3}$ and $l'_{3}$ the corresponding integer. The unit vector $\mathbf{e}_{z} = \mathbf{a}_{1} \times \mathbf{a}_{2}/ |\mathbf{a}_{1} \times \mathbf{a}_{2}| \parallel \mathbf{a}_{3} \parallel \mathbf{a}'_{3}$, clearly, it points from the upper electrode to the lower one, and thus antiparallel to the direction of the tunneling current. In equation \eqref{Vgamma}, we employ a delta potential well to describe approximately the effect from the interface, the parameter $\gamma$ being the strength. Because every atomic layer of the barrier (the moir\'{e} superlattice) is a 2D single crystal, and possesses a rotation symmetry of $60^{\circ}$ and a reflection symmetry of $30^{\circ}$, as can be seen in Fig. 1(c), the barrier will also possess the same symmetries as its layers, i.e., both a rotation symmetry of $60^{\circ}$ and a reflection symmetry of $30^{\circ}$. Evidently, the interface potential $V$ must be in accordance with the barrier in symmetry. That implies that the strength $\gamma$ must be a symmetric function of the twist angle $\alpha$. Heeding all those facts, we perform a Fourier expansion to the strength $\gamma$ and approximate it to the first order: $\gamma(\alpha) = \gamma_{0}[1 - \cos(6\alpha)]$ where $\gamma_{0}$ is the zeroth-order coeffiecient, which will be taken as a model paramer here (the first-order coeffiecient can be determined simply by the fact: $\gamma(0) =0$). To be more intuitive, the shape of function $\gamma(\alpha)$ is shown in Fig. 1(d) where both the rotation symmetry of $60^{\circ}$ and the reflection symmetry of $30^{\circ}$ become self-evident. As usual, we have added a very small imaginary part ($i\varepsilon$) to the interface potential $V$ so as to guarantee the convergence of the tunneling conductances \cite{rfb2,rfb3}. In the following numerical calculations, we shall set $\varepsilon = 10^{-8}$.

Now, suppose that an electron tunnels from the upper electrode into the lower one. The incident electron can be described by the plane wave as follows,
\begin{equation}
\psi _{i} = \exp (i\mathbf{k}\cdot \mathbf{r}),
\end{equation}
where $\mathbf{k}$ represents the wave vector, and $\mathbf{r}$ the position vector: $\mathbf{r} = x_{1} \mathbf{a}_{1} + x_{2} \mathbf{a}_{2} + z \mathbf{e}_{z}$ with $x_{1}$, $x_{2}$, and $z$ being the affine coordinates. According to the standard methods in diffraction physics, i.e. Bethe theory and two-beam approximation \cite{rf44,rf45,rf46,rf47,rf48}, the transmitted wave function $\psi(\mathbf{r})$ can be obtained as follows,
\begin{align}
\psi(\mathbf{r})=
&  \frac{1}{2} \Big[ A_{1}  \exp\Big(i\mathbf{k}^{++}\cdot(\mathbf{r}-d_{1}\mathbf{e}_{z})\Big)
 - A_{1} \exp\Big(i\mathbf{k}_{g}^{++}\cdot(\mathbf{r}-d_{1}\mathbf{e}_{z})\Big)
 - A_{2} \exp\Big(i\mathbf{k}_{h}^{++}\cdot(\mathbf{r}-d_{1}\mathbf{e}_{z})\Big)    \notag \\
&+ A_{2} \exp\Big(i\mathbf{k}_{m}^{++}\cdot(\mathbf{r}-d_{1}\mathbf{e}_{z})\Big)
 + A_{3} \exp\Big(i\mathbf{k}^{--}\cdot(\mathbf{r}-d_{1}\mathbf{e}_{z})\Big)
 + A_{3} \exp\Big(i\mathbf{k}_{g}^{--}\cdot(\mathbf{r}-d_{1}\mathbf{e}_{z})\Big)   \notag \\
&+ A_{4} \exp\Big(i\mathbf{k}_{h}^{--}\cdot(\mathbf{r}-d_{1}\mathbf{e}_{z})\Big)
 + A_{4} \exp\Big(i\mathbf{k}_{m}^{--}\cdot(\mathbf{r}-d_{1}\mathbf{e}_{z})\Big)
 + (A_{1} + A_{3}) \exp\Big(i\mathbf{k}\cdot(\mathbf{r}-d_{1}\mathbf{e}_{z})\Big)    \notag \\
&+ (A_{1} - A_{3}) \exp\Big(i\mathbf{p}_{g}\cdot(\mathbf{r}-d_{1}\mathbf{e}_{z})\Big)
 - (A_{2} - A_{4}) \exp\Big(i\mathbf{p}_{h}\cdot(\mathbf{r}-d_{1}\mathbf{e}_{z})\Big)  \notag \\
&- (A_{2} + A_{4})\exp\Big(i\mathbf{p}_{m}\cdot(\mathbf{r}-d_{1}\mathbf{e}_{z})\Big)\Big],
\end{align}
where
\begin{eqnarray}
  \mathbf{k}^{\pm\pm}
  &=&  \mathbf{k}_{\parallel} + \sqrt{\mathbf{k}^{2} - \mathbf{k}_{\parallel}^{2}\pm 2m\hbar ^{-2}\, v(\mathbf{K}_{h})
       \pm 2m\hbar ^{-2}\, v(\mathbf{K}_{g})} \mathbf{e}_{z},   \\
  \mathbf{k}_{h}^{\pm\pm}
  &=&  \mathbf{k}_{\parallel} + \mathbf{K}_{h}+\sqrt{\mathbf{k}^{2} - (\mathbf{k}_{\parallel} + \mathbf{K}_{h})^{2}
       \pm 2m\hbar ^{-2}\, v(\mathbf{K}_{h}) \pm 2m\hbar ^{-2}\, v(\mathbf{K}_{g})}\mathbf{e}_{z}, \\
  \mathbf{k}_{g}^{\pm\pm}
  &=&  \mathbf{k}_{\parallel} + \mathbf{K}_{g} + \sqrt{\mathbf{k}^{2} - (\mathbf{k}_{\parallel} + \mathbf{K}_{g})^{2}
       \pm 2m\hbar ^{-2}\, v(\mathbf{K}_{h}) \pm 2m\hbar ^{-2}\, v(\mathbf{K}_{g})}\mathbf{e}_{z},  \\
  \mathbf{k}_{m}^{\pm\pm}
  &=&  \mathbf{k}_{\parallel} + \mathbf{K}_{m} + \sqrt{\mathbf{k}^{2}-(\mathbf{k}_{\parallel} + \mathbf{K}_{m})^{2}
       \pm 2m\hbar ^{-2}\, v(\mathbf{K}_{h}) \pm 2m\hbar ^{-2}\, v(\mathbf{K}_{g})}\mathbf{e}_{z},  \\
  \mathbf{p}_{h}
  &=&  \mathbf{k}_{\parallel} + \mathbf{K}_{h}+\sqrt{\mathbf{k}^{2}
     - (\mathbf{k}_{\parallel} + \mathbf{K}_{h})^{2}}\mathbf{e}_{z}, \\
  \mathbf{p}_{g}
  &=&  \mathbf{k}_{\parallel} + \mathbf{K}_{g} + \sqrt{\mathbf{k}^{2}
     - (\mathbf{k}_{\parallel} + \mathbf{K}_{g})^{2}}\mathbf{e}_{z},  \\
  \mathbf{p}_{m}
  &=&  \mathbf{k}_{\parallel} + \mathbf{K}_{m} + \sqrt{\mathbf{k}^{2}
     -(\mathbf{k}_{\parallel} + \mathbf{K}_{m})^{2}}\mathbf{e}_{z},
\end{eqnarray}
and
\begin{eqnarray}
  A_{1}
  &=&  \frac{1}{2} \frac{\exp\left(i\mathbf{k}_{z}^{+}\cdot d_{1}\right)}
       {1+i\frac{m\gamma(\alpha)}{\hbar ^{2}\mathbf{k}_{z}^{+}}}, \quad
  A_{2}
  =    \frac{1}{2} \frac{\exp\left(i\mathbf{k}_{hz}^{+}\cdot d_{1}\right)}
       {1+i\frac{m\gamma(\alpha)}{\hbar ^{2}\mathbf{k}_{hz}^{+}}},\\
  A_{3}
  &=&  \frac{1}{2} \frac{\exp\left(i\mathbf{k}_{z}^{-}\cdot d_{1}\right)}
       {1+i\frac{m\gamma(\alpha)}{\hbar ^{2}\mathbf{k}_{z}^{-}}},  \quad
  A_{4}
  =    \frac{1}{2} \frac{\exp\left(i\mathbf{k}_{hz}^{-}\cdot d_{1}\right)}
       {1+i\frac{m\gamma(\alpha)}{\hbar ^{2}\mathbf{k}_{hz}^{-}}}. \label{A3A4}
\end{eqnarray}
Here, $\mathbf{k}_{\parallel}$ is the normal projection of the incident wave vector $\mathbf{k}$ on the plane spanned by $\mathbf{a}_{1}$ and $\mathbf{a}_{2}$. The $v(\mathbf{K}_{h})$ and $v(\mathbf{K}_{g})$ represent the Fourier transform of $v(\mathbf{r})$ where $\mathbf{K}_{h}$ and $\mathbf{K}_{g}$ are the planar vectors reciprocal to the intralayer lattice vectors $\mathbf{R}_{h}$ and $\mathbf{R}_{g}$, respectively. According to the two-beam approximation \cite{rf44,rf45,rf46,rf47,rf48}, one has
\begin{eqnarray}
  \mathbf{K}_{h} &=& (K,0,0), \\
  \mathbf{K}_{g} &=&
  \begin{cases}
    (K\cos(\alpha-\pi/3), -K\sin(\alpha-\pi/3), 0) & \varphi \in[-\pi/6,-\pi/6+\alpha]  \\
    (K\cos\alpha, -K\sin\alpha, 0) & \varphi \in[-\pi/6+\alpha,\pi/6],
  \end{cases} \label{Kgvalue} \\
  \mathbf{K}_{m} &=& (2K\sin^{2}(\alpha/2),2K\sin(\alpha/2)\cos(\alpha/2),0),  \label{Kmvalue}
\end{eqnarray}
where $K = 2\pi/a_{1}$ and $\varphi$ is the angle between $\mathbf{k}_{\parallel}$ and $\mathbf{a}_{1}$. Evidently, $\mathbf{K}_{h}$ and $\mathbf{K}_{g}$ differ from each other only by an angle $\alpha$ or $\pi/3-\alpha$. That implies $v(\mathbf{K}_{h}) = v(\mathbf{K}_{g})$. It can be seen that $\mathbf{K}_{m}$ is a member vector of the so-called mini Brillouin zone. At last,
\begin{eqnarray}
  \mathbf{k}_{z}^{\pm}
  &=&  \mathbf{k}_{\parallel} + \sqrt{\mathbf{k}^{2} - \mathbf{k}_{\parallel}^{2} - 2m\hbar ^{-2}\, v(0)
       \pm 2m\hbar ^{-2}\, v(\mathbf{K}_{h})} \mathbf{e}_{z},  \\
  \mathbf{k}_{hz}^{\pm}
  &=&  \mathbf{k}_{\parallel} + \mathbf{K}_{h} + \sqrt{\mathbf{k}^{2} - (\mathbf{k}_{\parallel}
     + \mathbf{K}_{h})^{2} - 2m\hbar ^{-2}\, v(0)\pm 2m\hbar ^{-2}\, v(\mathbf{K}_{h})} \mathbf{e}_{z},
\end{eqnarray}
where $v(0)$ is the zero-frequency component of the Fourier expansion of $v(\mathbf{r})$, i.e., the so-called average potential in the superlattice. The four are the $z$-components of wave vectors of the partial waves at the interface.

By using the out-going wave $\psi(\mathbf{r})$, the transmission coefficient can be calculated as follows \cite{rf44,rf45,rf46,rf47},
\begin{equation}\label{}
    T(\mathbf{k}) = \frac{|\mathbf{a}_{1} \times \mathbf{a}_{2}|}{2ik_{z}S_{h}}
    {\displaystyle\iint }\mathrm{d}x_{1}\mathrm{d}x_{2}\,\left[\psi^{\ast }(\mathbf{r})\,
    \frac{\partial}{\partial z} \psi(\mathbf{r}) - \mathrm{c.c.}\right],
\end{equation}
where $k_{z} = \mathbf{k}\cdot \mathbf{e}_{z}$, $S_{h}$ is the cross-sectional area of the barrier. From $T(\mathbf{k})$, the conductance $G$ of zero bias voltage at zero temperature can be written as \cite{rf44,rf45,rf46,rf47,rfb4}
\begin{equation}
G = \frac{e^{2}}{16\pi ^{3}\hbar }\int_{0}^{\pi /2}\mathrm{d}\theta \int_{0}^{2\pi }\mathrm{d}\varphi\, k_{F}^{2}\,
    \sin(2\theta)\,T\left(k_{F},\theta ,\varphi\right),
\end{equation}
where $e$ represents the electron charge, $\theta$ the angle between $\mathbf{k}$ and $\mathbf{e}_{z}$, and $k_{F}$ the Fermi wave vector of the tunneling electrons,
\begin{equation}
 k_{F}= \sqrt{\frac{2m}{\hbar^{2}} \mu}
\end{equation}
with $m$ being the electron mass, and $\mu$ the chemical potential of the electrodes. Here, for simplicity, we assume that the two electrodes are fabricated with the same metallic materials.

\section{Results and discussion}

From now on, we shall apply the above formalism to the case where the tunneling barrier is a graphite(graphene)/graphite(graphene) moir\'{e} superlattice. To this end, it needs totally five model parameters: the magnitude of the reciprocal-lattice vector $K$, the chemical potential $\mu$ of the electrodes, the Fourier transform of the periodic potential of the layers $v(\mathbf{K}_{h})$ ($v(\mathbf{K}_{h}) = v(\mathbf{K}_{g})$), the average potential $v(0)$, and the strength parameter $\gamma_{0}$ for the interface potential. From Ref. \cite{rf49}, $K = 4\pi/\sqrt{3}a_{1} = 2.95\times 10^{10} \,\mathrm{m}^{-1}$. According to the Fermi energy of Au \cite{rf50}, we set the chemical potential $\mu \approx 5.5\, \mathrm{eV}$. As to the other parameters, we set their values so as to agree with the experiments: $v(\mathbf{K}_{h}) \approx 3.5\, \mathrm{eV}$, $v(0) \approx 3.1\, \mathrm{eV}$, $\gamma_{0} = 4 \times 10^{-29}\, \mathrm{J\cdot m}$. In addition, following Ref. \cite{rf38}, we set $d_{1} + d_{2} = 50\, \mathrm{nm}$.

\subsection{The background and peak for the interlayer conductance}

First, we would like to invetigate the dependence of the interlayer (tunneling) conductance on the twist angle. The numerical results are displayed in Fig. 2 where $\mu = 5.5\, \mathrm{eV}$, $v(\mathbf{K}_{h(g)}) = 3.5\, \mathrm{eV}$, $v(0) = 3.12\, \mathrm{eV}$, $d_{1} = 1.78\, \mathrm{nm}$, and $d_{2} = 48.22\, \mathrm{nm}$. Because the interlayer conductance has a $30^{\circ}$ reflection symmetry and a $60^{\circ}$ periodicity, here, the twist angle only ranges from $0^{\circ}$ to $30^{\circ}$. Clearly, the interlayer conductance shows a decreasing background plus a peak around $21.8^{\circ}$ on the interval $[0^{\circ},\,30^{\circ}]$, which reproduces thoroughly the two most important features in experiments \cite{rf38,rf39}.

\begin{figure}[!ht]
\centering
\includegraphics[scale=0.35]{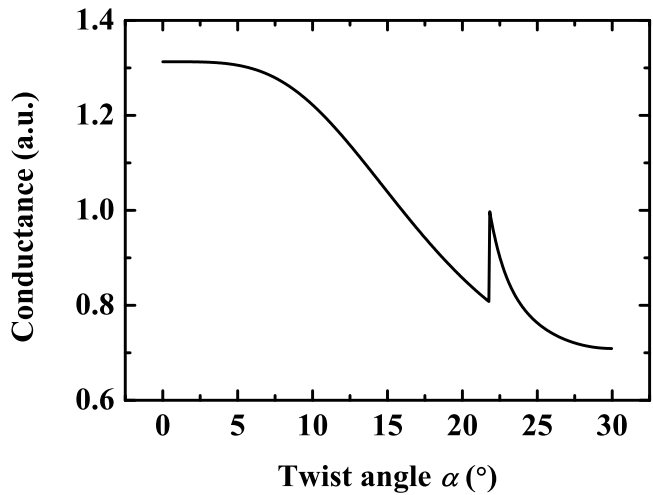}
\caption{Interlayer conductances as functions of twist angle $\alpha$.}
\end{figure}

Physically, there are totally two possible origins for the variation of the interlayer conductance with the twist angle $\alpha$: The first is the interface potential $V$. From equation \eqref{Vgamma}, we see that it is an explicit function of $\alpha$. As a scattering potential, it is natural that $V$ can directly result in the variation of interlayer conductance with the twist angle $\alpha$. The second is the vectors $\mathbf{K}_{g}$ and $\mathbf{K}_{m}$, they both are also functions of the twist angle $\alpha$, as indicated by equations \eqref{Kgvalue} and \eqref{Kmvalue}. That leads to that the wave vectors $\mathbf{k}_{g}^{\pm\pm}$,  $\mathbf{k}_{m}^{\pm\pm}$, $\mathbf{p}_{g}$, and $\mathbf{p}_{m}$ will all depend on $\alpha$. As a result, the transmitted wave function $\psi(\mathbf{r})$, the transmission coefficient $T(\mathbf{k})$ and the interlayer conductance $G$ will vary with the twist angle $\alpha$. Through numerical analysis, we find that the former is the main reason for the dependence of the interlayer conductance on the twist angle, the latter has little influence. In other words, the interface effect is just the basic mechanism for the variation of the interlayer conductance with the twist angle. That is in accordance with the viewpoint of Ref. \cite{rf39}. Since $\mathbf{K}_{m}$ belongs to mini Brillouin zone, it can also be inferred that the mini Brillouin zone, which is constructed from the difference between the reciprocal lattice vectors of the upper and lower single crytals, has little influence on the interlayer conductance. In the interval $[0^{\circ},\,30^{\circ}]$, the strength of the the interface potential $V$ is an increasing function of $\alpha$. Physically, the stronger the barrier potential, the weaker the transmission probability of the electrons except some isolated points where resonance may happen. Therefore, the interlayer conductance will decrease with the twist angle $\alpha$ if $\alpha \in [0^{\circ},\,30^{\circ}]$ except at resonance points. That accounts for the decreasing background of the interlayer conductance. On the other hand, if the twist angle $\alpha$ satisfies the condition that $1+i m\gamma(\alpha)/(\hbar^{2}\mathbf{k}_{z}^{-}) \rightarrow 0$, the amplitude $A_{3}$ becomes divergent, i.e., $|A_{3}| \rightarrow \infty$, as can be seen from equation \eqref{A3A4}, according to the quantum theory of scattering, a resonance will happen and form a peak at this twist angle to the interlayer conductance. That is the physical reason why some peaks can present in the interlayer conductance. From numerical calculations, we find that there can only emerge one peak around $\alpha = 21.8^{\circ}$ in the interval of $[0^{\circ},\,30^{\circ}]$ for appropriate parameters. So far, both the background and the peak have been interpreted in a unified way within the framework of the present theory.

\subsection{Effects of the physical parameters on the interlayer conductance}

Next, we would like to investigate the effects of the physical parameters on the interlayer conductances.

The curves of the interlayer conductance versus twist angles under different $d_{1}$ and $d_{2}$ are depicted in Fig. 3(a) and 3(b) respectively where $\mu = 5.5\, \mathrm{eV}$, $v(\mathbf{K}_{h(g)}) = 3.5\, \mathrm{eV}$, $v(0) = 3.12\, \mathrm{eV}$. In Fig. 3(a), $d_{2}$ is fixed as $48.22\, \mathrm{nm}$, and in Fig. 3(b), $d_{1}$ is fixed as $1.78\, \mathrm{nm}$. It can be seen from Fig. 3(a) that the peak intensity decreases significantly with a small increasing of $d_{1}$. This can be explained as follows: One can easily see that the $\mathbf{k}_{z}^{-}$ has an imaginary part around the resonance point. Therefore, the numerator $\exp(i\mathbf{k}_{z}^{-}\cdot d_{1})$ of the amplitude $A_{3}$ will attenuate exponentially with $d_{1}$ near the resonance point. Accordingly, the transmission coefficient $T$ and the interlayer conductance $G$ will also attenuate exponentially with the width $d_{1}$. This case of Fig. 3(a) is completely different from Fig. 3(b) where $d_{1}$ is fixed. Fig. 3(b) shows that the width $d_{2}$ has little effect on the peak intensity. That is because the amplitude $A_{3}$ is independent of $d_{2}$. As a result, the thickness of the upper single crystal of moir\'{e} superlattice is crucial for the presence of the conductance peak. From the present results, we can infer that the appearance of considerable conductance peaks needs a small $d_{1}$. If $d_{1}$ is larger than $10\, \mathrm{nm}$, the peaks can hardly be observed. This conclusion is in agreement with the Ref. \cite{rf39} where the upper single crystal is just a monolayer graphene.

\begin{figure}[!ht]
\centering
\includegraphics[scale=0.6]{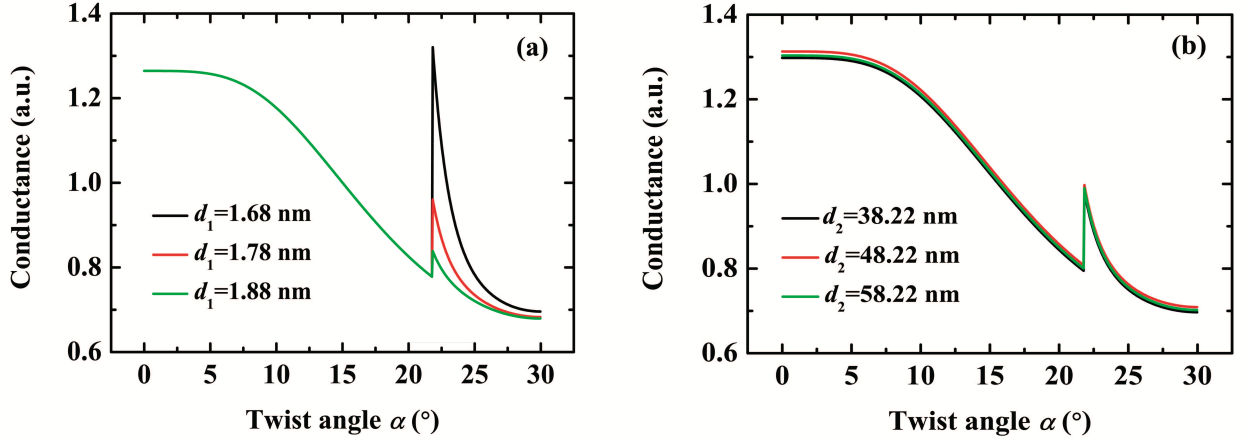}
\caption{Interlayer conductances as functions of twist angle $\alpha$ under different (a) $d_{1}$ and (b) $d_{2}$.}
\end{figure}

Fig. 4 shows the curves of the interlayer conductance versus twist angle under different chemical potential $\mu$ where $v(\mathbf{K}_{h(g)}) = 3.5\, \mathrm{eV}$, $v(0) = 3.12\, \mathrm{eV}$, $d_{1} = 1.78\, \mathrm{nm}$, and $d_{2} = 48.22\, \mathrm{nm}$. Obviously, both the intensity and position of the peak are affected explicitly by $\mu$: The intensity increases with $\mu$ whereas the position decreases with $\mu$. It is easy to know that the minimum of $|\mathbf{k}_{z}^{-}|$ will decrease with $\mu$. On one hand, such a property of $\mathbf{k}_{z}^{-}$ can make the numerator $\exp(i\mathbf{k}_{z}^{-}\cdot d_{1})$ of the amplitude $A_{3}$ increase with $\mu$, and thus enforce the peak intensity. On the other hand, it can also make the angle $\alpha$ corresponding to the resonance condition $1+i m\gamma(\alpha)/(\hbar^{2}\mathbf{k}_{z}^{-}) \rightarrow 0$ decrease with $\mu$, and thus shift the position of the peak towards the small twist angle. That explains qualitatively the results of Fig. 4. Here, the conclusion that the peak of the interlayer conductance will shift with the variation of the chemical potential of the electrodes is just a theoretical prediction, which needs the future verification of experiment.

\begin{figure}[!ht]
\centering
\includegraphics[scale=0.35]{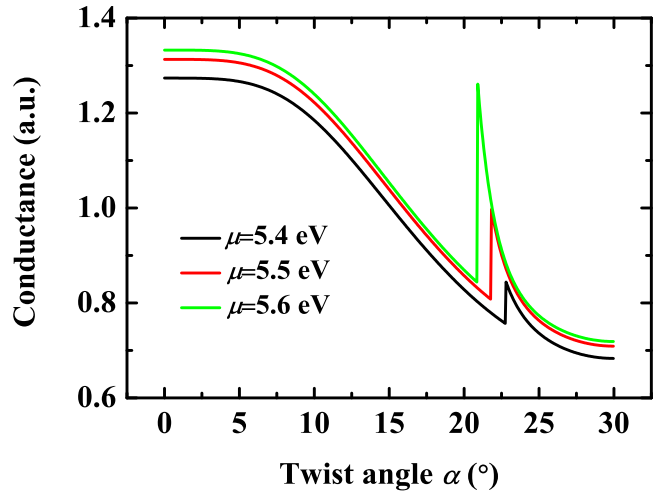}
\caption{Interlayer conductances as functions of twist angle $\alpha$ under different chemical potential $\mu$.}
\end{figure}

The influences of $v(\mathbf{K}_{h(g)})$ and $v(0)$ on the interlayer conductance are displayed in Fig. 5(a) and 5(b) respectively where $\mu = 5.5\, \mathrm{eV}$, $d_{1} = 1.78\, \mathrm{nm}$, and $d_{2} = 48.22\, \mathrm{nm}$. In Fig. 5(a), $v(0)$ is fixed as $3.12\, \mathrm{eV}$, and in Fig. 5(b), $v(\mathbf{K}_{h(g)})$ is fixed as $3.5\, \mathrm{eV}$. From Fig. 5, it can be found that the effects of $v(\mathbf{K}_{h(g)})$ and $v(0)$ on the conductance peak are similar, that is, the stronger the $v(\mathbf{K}_{h(g)})$ or $v(0)$, the weaker the intensity of the peak, and the large the twist angle of the peak. Similar to Fig. 4, those results of Fig. 5 are also due to the property that the minimum of $|\mathbf{k}_{z}^{-}|$ will increase with $v(\mathbf{K}_{h(g)})$ or $v(0)$. Like the case of Fig. 4, the conclusion here is also a theoretical prediction, and needs the future verification of experiment.

\begin{figure}[!ht]
\centering
\includegraphics[scale=0.6]{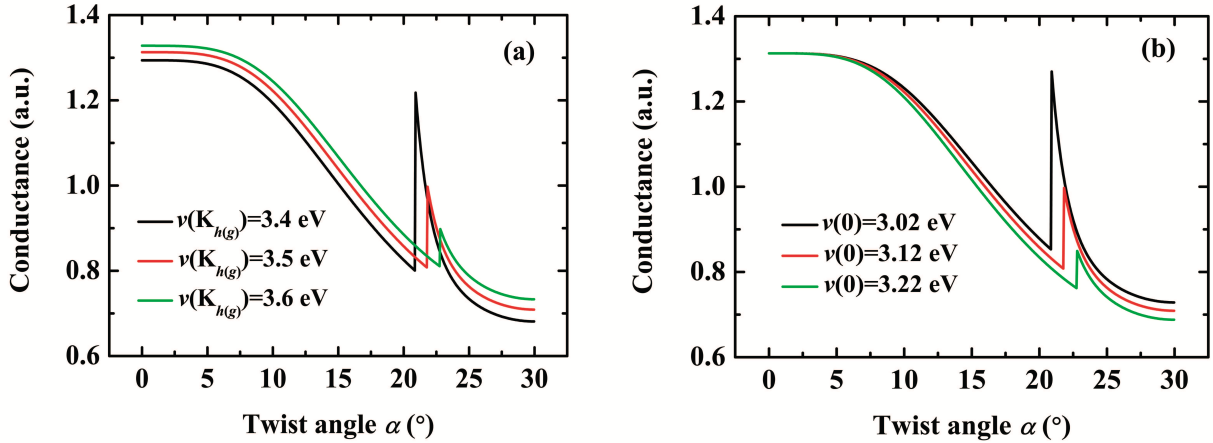}
\caption{Interlayer conductances as functions of twist angle $\alpha$ under different (a) $v(\mathbf{K}_{h(g)})$ and (b) $v(0)$.}
\end{figure}

To sum up, the intensity and position of the conductance peak can be regulated through the thickness of the upper single crystal as well as the materials of the electrodes and superlattices.

\subsection{A new type of electronic device: moir\'{e} tunnel junctions}

Finally, we need to discuss the topic from the view of device applications. Actually, as can be seen from Figs. 1(a)
and 1(b), the moir\'{e} superlattice plus the metallic electrodes on both sides forms a heterojunction with sandwich structure. Such heterojunctions have the advantages of simple configuration, reliable performance and rich physics, and thus are quite appropriate for using as basic units of devices. In the field of electronics, the heterojunctions with sandwich structure have been widely investigated and applied, such as MTJs. However, for the present case, the selection of the materials of electrodes and barrier, the physical structure of the barrier, and the working mechanism for the system are all quite different from MTJs. This indicates that the present system may become a new kind of electronic devices. Here, we shall call it ``moir\'{e} tunnel junctions". The existing experimental results have shown that moir\'{e} tunnel junctions possess great application potential. For example, according to the experiments and the present theoretical results, there are at least two kinds of physical mechanism for moir\'{e} tunnel junctions to realize the transformation between "open" and "close": (1) transform the twist angle between $0^{\circ}$ and $30^{\circ}$. (2) transform the twist angle between the peak and its vicinity. That is to say, moir\'{e} tunnel junctions, just like MTJs, can realize the transformation through the variation of tunneling current under different states. However, in MTJs, the variation of tunneling current is regulated through changing the relative direction of the magnetization of the electrodes on both sides, whereas, in moir\'{e} tunnel junctions, the variation of tunneling current is regulated through changing the twist angle between the two single crystals of the  moir\'{e} superlattices. In practical applications, MTJs have some disadvantages: First, there is strict limitation on the thickness of the barriers (ranging from several angstroms to a few nanometers). Secondly, it needs to impose the magnetic field and add the pinned layers to change the relative direction of the magnetization of the electrodes on both sides \cite{rf51}.  Delightfully, moir\'{e} tunnel junctions can overcome both the above disadvantages: The upper single crystal can be a single-layer 2D material, and the thickness of the lower single crystal can range from several angstroms to several tens of nanometers; it is unnecessary to impose any field or add additional layers, and the technique to regulate the twist angle through electric current has been mature \cite{rf38,rf42}. In order to see more clearly the differences between moir\'{e} tunnel junctions and MTJs, we list them in table 1. Of course, besides the above application, there are more broad developable prospects for moir\'{e} tunnel junctions, for example, the electrodes can be replaced by superconductors so as to develop superconducting devices.

\begin{table}
\renewcommand{\thetable}{\arabic{table}}
\newcommand{\tabincell}[2]{\begin{tabular}{@{}#1@{}}#2\end{tabular}}
  \centering
  \begin{tabular}{|c|c|c|c|c|c|}\hline
                    & \tabincell{c}{Moir\'{e} tunnel junctions} & \tabincell{c}{Magnetic tunnel junctions}  \\\hline
Electrode           & Non-magnetic metal                       &  Magnetic metal           \\\hline
Tunneling layer     & \tabincell{c}{The superlattice constructed by \\  the quasi-2D materials with moir\'{e} structure }    &  Insulated film           \\\hline
Physical mechanism  & \tabincell{c}{Regulate the tunneling current through the  \\  variation of twist angle of the moir\'{e} superlattices }          &  \tabincell{c}{Regulate the tunneling current through \\ changing the relative direction of the \\ magnetization of the electrodes on both sides}              \\\hline
\end{tabular}
  \caption{Comparison between moir\'{e} tunnel junctions and magnetic tunnel junctions}
\end{table}

\section{Conclusions}
In this paper, we have developed a microscopic theory for the interlayer conductance of moir\'{e} superlattice. Within the framework of the present theory, both the continuously decreasing background and the peak of the interlayer conductance observed in the experiments can be well explained by a unified mechanism. It is found that the interface potential plays a key role in the interlayer  conductance of moir\'{e} superlattice: The decreasing background arises from the increasing strength of the interface potential; the peak originates from the resonance of scattering of tunneling electrons with the interface potential. In addition, the theoretical results show that the thickness of the upper layered single crystal, the chemical potential of the electrodes, and the atomic potential of moir\'{e} superlattice can all have significant influences on the intensity and position of the peak whereas the thickness of the lower layered single crystal has not. Furthermore, we have proposed a new type electronic device, i.e. moir\'{e} tunnel junctions. In particular, the working mechanism for moir\'{e} tunnel junctions has been compared with magnetic tunnel junctions. It indicates that the moir\'{e} tunnel junctions have great potential in applications, and may become a new field in electronics.

\section{A\lowercase{uthor contributions}}
M. X. and H. F. conceived the project. H. F. carried out the calculations. M. X. led the analysis. H. F. and M. X. wrote the paper. M. X. supervised the study.

\section{N\lowercase{otes}}
The authors declare no competing interests.

\section{Acknowledgments}
This work is supported by the National Natural Science Foundation of China (11704197), the NUPTSF (NY217046).


\begin{thebibliography}{99}

\bibitem{rf1} Amidror, I. \textit{The Theory of the Moir\'{e} Phenomenon: Volumn \uppercase\expandafter{\romannumeral1}: Periodic Layers Ch. 1}; Springer-Verlag: London, 2009.

\bibitem{rf2} Rayleigh, L. On the manufacture and theory of diffraction-gratings. \textit{London, Edinburgh, and Dublin Philosophical Magazine}. \textbf{1874}, \textit{47}, 81-93.

\bibitem{rf3} Bistritzer, R.; MacDonald, A. H. Moir\'{e} bands in twisted double-layer graphene. \textit{Proc. Natl Acad. Sci. USA} \textbf{2011}, \textit{108}, 12233-12237.

\bibitem{rf4} Jin, C.; Regan, E. C.; Yan, A.; Utama, M. I. B.; Wang, D.; Zhao, S.; Qin, Y.; Yang, S.; Zheng, Z.; Shi, S.; Watanabe, K.; Taniguchi, T.; Tongay, S.; Zettl, A.; Wang, F. Observation of moir\'{e} excitons in WSe$_{2}$/WS$_{2}$ heterostructure superlattices. \textit{Nature} \textbf{2019}, \textit{567}, 76-80.

\bibitem{rf5} Seyler, K. L.; Rivera, P.; Yu, H.; Wilson, N. P.; Ray, E. L.; Mandrus, D. G.; Yan, J.; Yao, W.; Xu, X. Signatures of moir\'{e}-trapped valley excitons in MoS$_{2}$/WSe$_{2}$ heterobilayers. \textit{Nature} \textbf{2019}, \textit{567}, 66-70.

\bibitem{rf6} Yoo, H.; Engelke, R.; Carr, S.; Fang, S.; Zhang, K.; Cazeaux, P.; Sung, S. H.; Hovden, R.; Tsen, A. W.; Taniguchi, T.; Watanabe, T.; Yi, G. C.; Kim, M.; Luskin, M.; Tadmor, E. B.; Kaxiras, E.; Kim, P. Atomic and electronic reconstruction at the van der Waals interface in twisted bilayer graphene. \textit{Nat. Mater.} \textbf{2019}, \textit{18}, 448-453.

\bibitem{rf7} Tomarken, S. L.; Cao, Y.; Demir, A.; Watanabe, K.; Taniguchi, T.; Jarillo-Herrero, P.; Ashoori, R. C. Electronic compressibility of magic-angle graphene superlattices. \textit{Phys. Rev. Lett.} \textbf{2019}, \textit{123}, 046601.

\bibitem{rf8} Chen, P. Y.; Zhang, X. Q.; Lai, Y. Y.; Lin, E. C.; Chen, C. A.; Guan, S. Y.; Chen, J. J.; Yang, Z. H.; Tseng, Y. W.; Gwo, S.; Chang, C. S.; Chen, L. J.; Lee, Y. H. Tunable moir\'{e} superlattice of artificially twisted monolayers. \textit{Adv. Mater.} \textbf{2019}, \textit{31}, 1901077.

\bibitem{rf9} Xie, Y.; Lian, B.; J\"{a}ck, B.; Liu, X.; Chiu, C. L.; Watanabe, K.; Taniguchi, T.; Bernevig, B. A.; Yazdani, A. Spectroscopic signatures of many-body correlations in magic-angle twisted bilayer graphene. \textit{Nature} \textbf{2019572}, \textit{572}, 101-105.

\bibitem{rf10} Chen, G.; Sharpe, A. L.; Gallagher, P.; Rosen, I. T.; Fox, E. J.; Jiang, L.; Lyu, B.; Li, H.; Watanabe, K.; Taniguchi, T.; Jung, J.; Shi, Z.; Goldhaber-Gordon, D.; Zhang, Y.; Wang, F. Signatures of tunable superconductivity in a trilayer graphene moir\'{e} superlattice. \textit{Nature} \textbf{2019}, \textit{572}, 215-219.

\bibitem{rf11} Yan, C.; Ma, D. L.; Qiao, J. B.; Zhong, H. Y.; Yang, L.; Li, S. Y.; Fu, Z. Q.; Zhang, Y.; He, L. Scanning tunneling microscopy study of the quasicrystalline $30^{\circ}$ twisted bilayer graphene. \textit{2D Mater.} \textbf{2019}, \textit{6}, 045041.

\bibitem{rf12} Kim, K.; Dasilva, A.; Huang, S.; Fallahazad, B.; Larentis, S.; Taniguchi, T.; Wanatabe, K.; LeRoy, B. J.; MacDonald, A. H.; Tutuc, E. Tunable moir\'{e} bands and strong correlations in small-twist-angle bilayer graphene. \textit{Proc. Natl Acad. Sci. USA} \textbf{2017}, \textit{114}, 3364-3369.

\bibitem{rf13} Cao, Y.; Fatemi, V.; Demir, A.; Fang, S.; Tomarken, S. L.; Luo, J. Y.; Sanchez-Yamagishi, J. D.; Watanabe, K.; Taniguchi, T.; Kaxiras, E.; Ashoori, R. C.; Herrero, P. J. Correlated insulator behaviour at half-filling in magic-angle graphene superlattices. \textit{Nature} \textbf{2018}, \textit{556}, 80-84.

\bibitem{rf14} Cao, Y.; Fatemi, V.; Fang, S.; Watanabe, K.; Taniguchi, T.; Kaxiras, E.; Herrero, P. J. Unconventional superconductivity in magic-angle graphene superlattices. \textit{Nature} \textbf{2018}, \textit{556}, 43-50.

\bibitem{rf15} Huder, L.; Artaud, A.; Quang, T. L.; de Laissardi\`{e}re, G. T.; Jansen, A. G. M.; Lapertot, G.; Chapelier, C.; Renard, V. T. Electronic spectrum of twisted graphene layers under heterostrain. \textit{Phys. Rev. Lett.} \textbf{2018}, \textit{120}, 156405.

\bibitem{rf16} Chung, T. F.; Xu, Y.; Chen, Y. P. Transport measurements in twisted bilayer graphene: Electron-phonon coupling and Landau level crossing. \textit{Phys. Rev. B} \textbf{2018}, \textit{98}, 035425.

\bibitem{rf17} Marchenko, L.; Evtushinsky, D. V.; Golias, E.; Varykhalov, A.; Seyller, Th.; Rader, O. Extremely flat band in bilayer graphene. \textit{Sci. Adv.} \textbf{2018}, \textit{4}, eaau0059.

\bibitem{rf18} Qiao, J. B.; Yin, L. J.; He, L. Twisted graphene bilayer around the first magic angle engineered by heterostrain. \textit{Phys. Rev. B} \textbf{2018}, \textit{98}, 235402.

\bibitem{rf19} Yankowitz, M.; Chen, S.; Polshyn, H.; Zhang, Y.; Watanabe, K.; Taniguchi, T.; Graf, D.; Young, A. F.; Dean, C. R. Tuning superconductivity in twisted bilayer graphene. \textit{Science} \textbf{2019}, \textit{363}, 1059-1064.

\bibitem{rf20} Patel, H.; Huang, L.; Kim, C. J.; Park, J.; Graham M. W. Stacking angle-tunable photoluminescence from interlayer exciton states in twisted bilayer graphene. \textit{Nat. Commun.} \textbf{2019}, \textit{10}, 1445.

\bibitem{rf21} Liu, Y. W.; Qiao, J. B.; Yan, C.; Zhang, Y.; Li, S. Y.; He, L. Magnetism near half-filling of a Van Hove singularity in twisted graphene bilayer. \textit{Phys. Rev. B} \textbf{2019}, \textit{99}, 201408(R).

\bibitem{rf22} Kerelsky, A.; McGilly, L. J.; Kennes, D. M.; Xian, L.; Yankowitz, M.; Chen, S.; Watanabe, K.; Taniguchi, T.; Hone, J.; Dean, C.; Rubio, A.; Pasupathy, A. N. Maximized electron interactions at the magic angle in twisted bilayer graphene. \textit{Nature} \textbf{2019}, \textit{572}, 95-100.

\bibitem{rf23} Sharpe, A. L.; Fox, E. J.; Barnard, A. W.; Finney, J.; Watanabe, K.; Taniguchi, T.; Kastner, M. A.; Goldhaber-Gordon, D. Emergent ferromagnetism near three-quarters filling in twisted bilayer graphene. \textit{Science} \textbf{2019}, \textit{365}, 605-608.

\bibitem{rf24} Xu, S. G.; Berdyugin, A. I.; Kumaravadivel, P.; Guinea, F.; Kumar, R. K.; Bandurin, D. A.; Morozov, S. V.; Kuang, W.; Tsim, B.; Liu, S.; Edgar, J. H.; Grigorieva, I. V.; Falko, V. I.; Kim, M.; Geim, A. K. Giant oscillations in a triangular network of one-dimensional states in marginally twisted graphene. \textit{Nat. Commun.} \textbf{2019}, \textit{10}, 4008.

\bibitem{rf25} Jiang, Y.; Lai, X.; Watanabe, K.; Taniguchi, T.; Haule, K.; Mao, J.; Andrei, E. Y. Charge order and broken rotational symmetry in magic-angle twisted bilayer graphene. \textit{Nature} \textbf{2019}, \textit{573}, 91-95.

\bibitem{rf26} Guo, H.; Wang, X.; Lu, H.; Bao, L.; Peng, H.; Qian, K.; Ma, J.; Li, G.; Huang, L.; Lin, X.; Zhang, Y. Y.; Du, S.; Pantelides, S. T.; Gao, H. J. Centimeter-scale, single-crystalline, AB-stacked bilayer graphene on insulating substrates. \textit{2D Mater.} \textbf{2019}, \textit{6}, 045044.

\bibitem{rf27} Zhang, X.; Zhang, R.; Wang, Y.; Zhang, Y.; Jiang, T.; Deng, C.; Zhang, X.; Qin, S. In-plane anisotropy in twisted bilayer graphene probed by Raman spectroscopy. \textit{Nanotechnology} \textbf{2019}, \textit{30}, 435702.

\bibitem{rf28} Kulothungan, J.; Muruganathan, M.; Mizuta, H. Modulation of twisted bilayer CVD graphene interlayer resistivity by an order of magnitude based on in-situ annealing. \textit{Carbon} \textbf{2019}, \textit{153}, 355-363.

\bibitem{rf29} Polshyn, H.; Yankowitz, M.; Chen, S.; Zhang, Y.; Watanabe, K.; Taniguchi, T.; Dean, C. R.; Young, A. F. Large linear-in-temperature resistivity in twisted bilayer graphene. \textit{Nat. Phys.} \textbf{2019}, \textit{15}, 1011-1016.

\bibitem{rf30} Li, Z.; Zhuang, J.; Wang, L.; Feng, H.; Gao, Q.; Xu, X.; Hao, W.; Wang, X.; Zhang, C.; Wu, K.; Dou, S. X.; Chen, L.; Hu, Z.; Du, Y. Realization of flat band with possible nontrivial topology in electronic Kagome lattice. \textit{Sci. Adv.} \textbf{2018}, \textit{4}, eaau4511.

\bibitem{rf31} Zhang, X.; Zhang, R.; Zhang, Y.; Jiang, T.; Deng, C.; Zhang, X.; Qin, S. Tunable photoluminescence of bilayer MoS$_{2}$ via interlayer twist. \textit{Opt. Mater.} \textbf{2019}, \textit{94}, 213-216.

\bibitem{rf32} Kim, K.; Prasad, N.; Movva, H. C. P.; Burg, G. W.; Wang, Y.; Larentis, S.; Taniguchi, T.; Watanabe, K.; Register, L. F.; Tutuc, E. Spin-conserving resonant tunneling in twist-controlled WSe$_{2}$-hBN-WSe$_{2}$ heterostructures. \textit{Nano Lett.} \textbf{2018}, \textit{18}, 5967-5973.

\bibitem{rf33} Zhang, Q.; Yu, J.; Ebert, P.; Zhang, C.; Pan, C. R.; Chou, M. Y.; Shih, C. K.; Zeng, C.; Yuan, S. Tuning band gap and work function modulations in monolayer hBN/Cu(111) heterostructures with moir\'{e} patterns. \textit{ACS Nano} \textbf{2018}, \textit{12}, 9355-9362.

\bibitem{rf34} Sanctis, A. D.; Mehew, J. D.; Alkhalifa, S.; Withers, F.; Craciun, M. F.; Russo, S. Strain-engineering of twist-angle in graphene/hBN superlattice devices. \textit{Nano Lett.} \textbf{2018}, \textit{18}, 7919-7926.

\bibitem{rf35} Merkl, P.; Mooshammer, F.; Steinleitner, P.; Girnghuber, A.; Lin, K. Q.; Nagler, P.; Holler, J.; Sch\"{u}ller, C.; Lupton, J. M.; Korn, T.; Ovesen, S.; Brem, S.; Malic, E.; Huber, R. Ultrafast transition between exciton phases in van der Waals heterostructures. \textit{Nat. Mater.} \textbf{2019}, \textit{18}, 691-696.

\bibitem{rf36} Gogoi, P. K.; Lin, Y. C.; Senga, R.; Komsa, H. P.; Wong, S. L.; Chi, D.; Krasheninnikov, A. V.; Li, L. J.; Breese, M. B. H.; Pennycook, S. J.; Wee, A. T. S.; Suenaga, K. Layer rotation-angle-dependent excitonic absorption in van der Waals heterostructures revealed by electron energy loss spectroscopy. \textit{ACS Nano} \textbf{2019}, \textit{13}, 9541-9550.

\bibitem{rf37} Yu, Z.; Song, A.; Sun, L.; Li, Y.; Gao, L.; Peng, H.; Ma, T.; Liu, Z.; Luo, J. Understanding interlayer contact conductance in twisted bilayer graphene. \textit{Small} \textbf{2019}, \textit{16}, 1902844.

\bibitem{rf38} Koren, E.; Leven, I.; L\"{o}rtscher, E.; Knoll, A.; Hod, O.; Duerig, U. Coherent commensurate electronic states at the interface between misoriented graphene layers. \textit{Nat. Nanotechnol.} \textbf{2016}, \textit{11}, 752-757.

\bibitem{rf39} Chari, T.; Ribeiro-Palau, R.; Dean, C. R.; Shepard, K. Resistivity of rotated graphite-graphene contacts. \textit{Nano Lett.} \textbf{2016}, \textit{16}, 4477-4482.

\bibitem{rf40} Liao, M.; Wu, Z. W.; Du, L.; Zhang, T.; Wei, Z.; Zhu, J.; Yu, H.; Tang, J.; Gu, L.; Xing, Y.; Yang, R.; Shi, D.; Yao, Y.; Zhang, G. Twist angle-dependent conductivities across MoS$_{2}$/graphene heterojunctions. \textit{Nat. Commun.} \textbf{2018}, \textit{9}, 4068.

\bibitem{rf41} Kim, Y.; Yun, H.; Nam, S. G.; Son, M.; Lee, D. S.; Kim, D. C.; Seo, S.; Choi, H. C.; Lee, H. J.; Lee, S. W.; Kim, J. S. Breakdown of the interlayer coherence in twisted bilayer graphene. \textit{Phys. Rev. Lett.} \textbf{2013}, \textit{110}, 096602.

\bibitem{rf42} Li, H.; Wei, X.; Wu, G.; Gao, S.; Chen, Q.; Peng, L. M. Interlayer electrical resistivity of rotated graphene layers studied by in-situ scanning electron microscopy. \textit{Ultramicroscopy} \textbf{2018}, \textit{193}, 90-96.

\bibitem{rf43} Bistritzer, R.; MacDonald, A. H. Transport between twisted graphene layers. \textit{Phys. Rev. B} \textbf{2010}, \textit{81}, 245412.

\bibitem{rf44} Fang, H. N.; Xiao, M. W.; Rui, W. B.; Du, J.; Tao, Z. K. Magnetic coherent tunnel junctions with periodic grating barrier. \textit{Sci. Rep.} \textbf{2016}, \textit{6}, 24300.

\bibitem{rf45} Fang, H. N.; Xiao, M. W.; Zhong, Y. Y.; Rui, W. B.; Du, J.; Tao, Z. K. Magnetic tunnel junctions consisting of a periodic grating barrier and two half-metallic electrodes. \textit{New J. Phys.} \textbf{2019}, \textit{21}, 123006.

\bibitem{rf46} Fang, H. N.; Xiao, M. W.; Rui, W. B.; Du, J.; Tao, Z. K. K. Effects of temperature on the magnetic tunnel junctions with periodic grating barrier. \textit{J. Magn. Magn. Mater.} \textbf{2018}, \textit{465}, 333-338.

\bibitem{rf47} Fang, H. N.; Zang, X.; Xiao, M. W.; Zhong, Y. Y.; Tao, Z. K. Oscillations of tunneling magnetoresistance on bias voltage in magnetic tunnel junctions with periodic grating barrier. \textit{J. Appl. Phys.} \textbf{2020}, \textit{127}, 163902.

\bibitem{rf48} Cowley, J. M. \textit{Diffraction Physics}; Elsevier: Amsterdam-Lausanne-New York-Oxford-Shannon-Tokyo, 1995.

\bibitem{rf49} Chung, D. D. L. Review graphite. \textit{J. Mater. Sci.} \textbf{2002}, \textit{37}, 1475-1489.

\bibitem{rfb2} Molin\`{a}s-Mata, P.; Molin\`{a}s-Mata, P. Electron absorption by complex potentials: One-dimensional case. \textit{Phys. Rev. A} \textbf{1996}, \textit{54}, 2060-2065.

\bibitem{rfb3} Ko\v{c}inac, S. L. S.; Milanovi\'{c}, V. Tunneling times in complex potentials. \textit{Phys. Lett. A} \textbf{2008}, \textit{372}, 191-196.

\bibitem{rfb4} Slonczewski, J. C. Conductance and exchange coupling of two ferromagnets seperated by a tunneling barrier. \textit{Phys. Rev. B} \textbf{1989}, \textit{39}, 6995-7002.

\bibitem{rf50} Ashcroft, N. W.; Mermin, N. D. \textit{Solid State Physics}; Cengage Learning, Inc, 1976.

\bibitem{rf51} Zhu, J.; Park, C. Magnetic tunnel junctions. \textit{Mater. Today} \textbf{2006}, \textit{9}, 36-45.

\end{thebibliography}
\end{document}